\documentclass[9pt]{pnas-new}
\usepackage[font=small,labelfont=bf,justification=justified,calcwidth =.99\linewidth]{caption}
\usepackage{xcolor}
\usepackage{amsmath,amssymb}
\usepackage[separate-uncertainty,detect-family=true,mode=math]{siunitx}
\usepackage{bm}

\renewcommand*{\vec}[1]{\bm{#1}}

\newcommand*{\tnsr}[1]{\bm{\mathsf{#1}}}  
\newcommand*{\norm}[1]{\lVert #1 \rVert}
\newcommand*{\avg}[1]{\left\langle #1 \right\rangle}
\newcommand*{\diff}{\mathop{}\!\mathrm{d}}

\usepackage{natbib}

\makeatletter
\renewcommand{\maketitle}{\noindent\textbf{\Huge\@title}\bigskip\par{\@author}\bigskip\par}
\makeatother

\title{Interfacial flow around a pusher bacterium}
\author{Jiayi Deng, Mehdi Molaei, Nicholas G. Chisholm and Kathleen J. Stebe}
\correspondingauthor{Kathleen J. Stebe\\E-mail: kstebe@seas.upenn.edu}

\begin{document}

\maketitle

Motile bacteria play essential roles in biology that rely on their dynamic behaviors, including their ability to navigate, interact, and self-organize. However, bacteria dynamics on fluid interfaces are not well understood. Swimmers adsorbed on fluid interfaces remain highly motile, and fluid interfaces are highly non-ideal domains that alter swimming behavior. To understand these effects, we study flow fields generated by \textit{Pseudomonas aeruginosa}
PA01 in the pusher mode. Analysis of correlated displacements of tracers and bacteria reveals dipolar flow fields with unexpected asymmetries that differ significantly from their counterparts in bulk fluids. We decompose the flow field into fundamental hydrodynamic modes for swimmers in incompressible fluid interfaces. We find an expected force-doublet mode corresponding to propulsion and drag at the interface plane, and a second dipolar mode, associated with forces exerted by the flagellum on the cell body in the aqueous phase that are countered by Marangoni stresses in the interface. The balance of these modes depends on the bacteria’s trapped interfacial configurations. Understanding these flows is broadly important in nature and in the design of biomimetic swimmers.

%

\section{Introduction}
\label{sec:intro}

Bacteria motility plays an essential role in bacteria survival. For example, the ability to generate flow or to self-propel allows motile bacteria to navigate, reorient and interact collectively. The remarkable ability of swimming bacteria to move at rates of up to 10 body lengths per second has generated intense interest in understanding their propulsion mechanisms. Furthermore, by understanding the motion and flow generated by biological swimmers, one can develop design rules for artificial biomimetic systems to recapitulate such behavior \citep{kokot_active_2017,spellings_shape_2015,sokolov_swimming_2010,shklarsh_smart_2011,di_leonardo_bacterial_2010}.
 
Swimming bacteria share a common and simple machinery, with rotating motors in their cell envelope that are coupled to the flagellum enabling bacterial self-propulsion \citep{terashima_chapter_2008}.
They typically move in Stokes flow, with negligible inertia. The propulsive force generated by rotation of their helical flagella, which drives their translational motion, is balanced by viscous drag. Thus, the leading-order flow field around an immersed swimmer is described by a force dipole or stresslet with ``pusher" or ``puller" modes depending on the dipole polarity \citep{lauga_hydrodynamics_2009}. In bulk fluids, the force dipole captures the main features of far-field flows generated by natural and artificial swimmers \citep{drescher_fluid_2011,campbell_experimental_2019,drescher_direct_2010}. Because of their directed motion, swimmers accumulate near boundaries \citep{lauga_hydrodynamics_2009,di_leonardo_swimming_2011,giacche_hydrodynamic_2010,lopez_dynamics_2014}. Hydrodynamic trapping near solid surfaces generates curvilinear trajectories \citep{lauga_swimming_2006,molaei_failed_2014, molaei_imaging_2014}; such trapping can promote surface colonization.

While bacteria hydrodynamics in the bulk and near solid boundaries has been extensively studied, their dynamics on or near fluid interfaces are less understood.
Motile bacteria are known to accumulate at interfaces between immiscible fluids \citep{bianchi_3d_2019,ahmadzadegan_hydrodynamic_2019,desai_hydrodynamics-mediated_2018,lopez_dynamics_2014} with curvilinear trajectories attributed to asymmetric drag in the vicinity of the interface \citep{deng_motile_2020,lemelle_counterclockwise_2010,morse_molecular_2013,pimponi_hydrodynamics_2016,lemelle_curvature_2013}. However, interfaces have other distinct features whose impact on swimmer dynamics are unexplored. Interfacial tension favors the adsorption of colloidal scale objects. Once adsorbed, these objects are trapped by significant energy barriers to desorption \citep{pieranski_two-dimensional_1980}, often with pinned contact lines that evolve slowly toward an equilibrium position \citep{kaz_physical_2012}. Furthermore, interfaces can have complex surface stresses including surface viscosities and Marangoni stresses owing to surfactant adsorption. These effects are associated with anomalous drag \citep{dani_hydrodynamics_2015,das_pairwise_2021,fischer_drag_2004,villa_motion_2020,pozrikidis_particle_2007,dorr_drag_2016} and divergence-free interfacial flow   \citep{desai_biofilms_2020,chisholm_driven_2021,fischer_viscous_2006,blawzdziewicz_stokes_1999}. While such factors dramatically restructure flow around passive colloidal particles at interfaces \citep{molaei_interfacial_2021}, their impact on the flow field generated by swimmers is unknown. 

In our previous work, we studied the behavior of the monotrichous bacteria \textit{P. aeruginosa} at the interface of hexadecane and an aqueous suspension of bacteria.  The bacteria were trapped with pinned contact lines at the fluid interface in a variety of configurations, with four categories of swimming trajectories which were characterized in terms of their statistical properties. The observed trajectory types include visitors that come and go from the interfacial plane; Brownian trajectories with bacteria diffusivities similar to those of inert colloidal particles; pirouettes for which bacteria spin furiously but move with diffusivities like those of inert colloids; and curly trajectories, for which bacteria swim along curly paths with trajectory curvature $\kappa$ ranging from \SI{1}{\micro m} to \SI{10}{\micro m} and swimming speeds up to \SI{40}{\micro m/s}. The curly trajectories, which were the most prevalent, were generated by bacteria swimming in pusher and puller modes. Clockwise (CW) segments of bacteria trajectories viewed from the water phase correspond to pusher motion \citep{deng_motile_2020}, whereas counterclockwise (CCW) segments correspond to swimming in puller mode. Switching between modes was achieved by reversal of the sense of rotation of a bacterium's single flagellum.  In this paper, we characterize the flow field generated by the bacteria in pusher mode.

In this work, we study the far-field, time-averaged flow generated by bacteria at fluid interfaces in the pusher mode and find that it deviates significantly from the typical stresslet observed for swimmers in bulk or near solid surfaces \citep{drescher_fluid_2011}, with a striking lack of fore-aft symmetry.
This flow field can be described analytically by two hydrodynamic dipolar modes.
One mode is associated with balancing thrust and drag forces parallel to the interfacial plane and is conceptually similar to the usual stresslet flow known to be produced by swimmers in bulk fluids. However, the observed flow is altered significantly from the bulk fluid case due to Marangoni effects. These effects dominate even when only trace surfactant is adsorbed  on the interface and force the interface to act as an incompressible layer.
The other mode is associated with stresses exerted on the fluid above and/or below the interface due to the motion of the swimming bacterium.
These off-interface, bulk stresses generate a purely Marangoni-driven mode of flow that is again associated with interfacial incompressibility. This latter mode has no analog for swimmers completely immersed in the bulk fluid. 

This paper is divided into seven sections, of which this Introduction is the first. In Section \ref{sec:exp_method}, we describe our experimental methods, which allow us to accurately approximate the flow disturbances generated by individual bacteria adsorbed to a fluid interface. In Section \ref{viscosity}, we describe the physico-chemistry of the fluid interface and its implications in constraining the flow generated by the swimmers. In section \ref {trap angle} we describe the characterization of the trapped configuration of cell bodies in the fluid interface. In Section \ref {sec:hydrodynamic-theory} we describe the interfacial flow generated by swimmers moving in pusher mode at the interface, and compare our experimental findings to prediction of fundamental hydrodynamic modes for swimmers in incompressible fluid interfaces. In Sections \ref{sec:discussion} and \ref{sec:conclusion}, we discuss the implications of our results and the conclusions from this study.

\section{Materials and Methods}
\label{sec:exp_method}

\subsection{Observation of bacteria trajectories}
\label{sec:materials}

\textit{P. aeruginosa} PA01 wild type bacteria were cultured and resuspended with Tris-based motility buffer. 
Swimming bacteria were observed at the interface formed by layering an aqueous bacteria suspension and hexadecane formed within a cylindrical vessel with inner diameter of \SI{1}{cm}. The cylindrical vessel's bottom half was made of aluminum, and its top half was made of Teflon. The vessel is filled with aqueous bacteria suspension (\SI{\sim 115}{\micro\liter}) to the aluminum-Teflon seam to form a planar interface. A hexadecane layer ($\sim$ \SI{100} to \SI{500} {\micro\liter}) was then placed over the aqueous suspension. 
The bottom surface of the cylinder is a glass coverslip, allowing the interface to be imaged from below. The bacterial suspension was sufficiently dilute that its viscosity was assumed to be that of water, \SI{0.89}{cP}, while the known viscosity of  hexadecane is \SI{3.45}{cP}.
A drop of \SI{0.5}{\micro m} diameter polystyrene tracer particles in isopropyl alcohol was spread on the planar aqueous-hexadecane interface.

Bacteria and tracer particles adsorbed on the interface were observed under an inverted microscope with a $40\times$ air objective (NA:0.55).
Video of the interface with area of \SI{217.2}{\micro m} by \SI{289.2}{\micro m} and pixel size of \SI{0.15}{\micro m} was recorded at \SI{25}{fps} for \SI{10}{min} with CMOS camera (Point Grey).
A typical number density of bacteria and tracers on the interface \SI{\sim 200}{mm^{-2}}, Figure~\ref{fig:CDV}(a).
Trajectories of the bacteria and tracers were tracked over time, Figure~\ref{fig:CDV}(b).
We address only bacteria that were swimming in clockwise circles in our results.
These bacteria swam forward (with cell body in front of flagellum) and are ``pushers'' or extensile swimmers.
The pusher trajectories are studied in detail to reveal the flow field generated by interfacially trapped swimmers moving in the pusher mode.
There is also a population of CCW swimmers that operate in reverse, in a ``puller'' or contractile swimming mode.
We exclude this sub-population from consideration for reasons detailed in Sec.~\ref{sec:discussion}.

\subsection{Correlated Displacement Velocimetry}
\label{sec:CDV-theory}

Correlated displacement velocimetry (CDV) was introduced by \cite{molaei_interfacial_2021} for measuring flow disturbances associated with the Brownian motion of colloidal particles.
Here, we describe this technique in detail and adapt it to measure the far-field flow field generated by an actively swimming bacterium.
CDV relies on the fact that bacteria move in an uncorrelated manner in a sufficiently dilute interfacial suspension of swimmers and passive tracer particles.
Therefore, we can statistically isolate the approximate flow disturbance due to the locomotion of an individual bacterium by correlating the displacements of each bacterium with the displacements of the surrounding tracer particles over a short time interval or ``lag time'' and superposing the results.
Specifically, we obtain the flow field by calculating the covariance of the displacement between pairs of swimmers and tracers.
This process filters out the effects of thermal diffusion of the tracer particles and also filters the effects of swimmer-swimmer and tracer-tracer interactions.
Moreover, CDV allows us to employ relatively dense tracer particles and swimmers in comparison to standard particle velocimetry techniques.

Here, we show how the covariance of swimmers and tracer displacements over a short period of time reveals the velocity field induced by the swimmer.
Consider swimmers interacting with a field of passive tracer particles.
Swimmers and tracers are assumed to be adsorbed on the interface and are therefore restricted to translate on the interfacial plane.
The position of the tracer particle and swimmer are denoted \(\vec x_{\rm p}\) and \(\vec x_{\rm s}\) (or \(\vec x_{\rm s}^n\) for the \(n\)th swimmer), respectively; these vectors are measured from an arbitrary origin in the laboratory frame.
The displacements of bacteria and tracers are treated as random variables.
Their correlated motion is determined by measuring the covariance of their displacements over the observation lag time.
During lag time \(\tau\), the \(n\)th swimmer undergoes displacement \(\Delta \vec x_{\rm s}^n(\tau)\) and a passive tracer located at position \(\vec x\) undergoes displacement \(\Delta \vec x_{\rm p}(\vec x, \tau)\).

Considering first the motion of the tracer particles, the displacement of each passive tracer has contributions from the flow field induced by all swimmers in the system.
Tracers also undergo Brownian motion and may interact with one another through electrostatic, capillary, or other means.
Thus, the displacement vector of a given tracer particle centered at position \(\vec x\) over lag time \(\tau\) interacting with \(N_{\rm s}\) swimmers and \(N_{\rm p}-1\) other particles can be expressed as
\begin{equation}
  \label{eq:particle-displacement}
  \Delta \vec x_{\rm p}(\vec x, \tau) = \sum_{n=1}^{N_{\rm s}}{\Delta\vec{x}_{\rm sp}^{n}}(\vec x, \tau) +
  \Delta\vec{x}_{\rm p,T}(\vec x, \tau) +
  \sum_{n=1}^{N_{\rm p} - 1}{\Delta\vec x_{\rm pp}^{n}}(\vec x, \tau),
\end{equation}
where \(\Delta\vec x_{\rm sp}^{n}(\vec x, \tau)\) is the displacement of a particle at \(\vec x\) induced by the flow disturbance of the \(n\)th swimmer, \(\Delta\vec x_{\rm p,T}\) is the Brownian displacement of the tracer (which has zero mean value), and \(\Delta\vec x_{\rm pp}^{n}\) represents the displacement due to interactions with the \(n\)th (other) tracer particle.

The displacement of the swimmer during lag time \(\tau\) is given by \(\Delta \vec x_{\rm s}(\tau) = \int_0^\tau \vec v_{\rm s}(t) \diff t\), where \(\vec v_{\rm s}\) is the swimming velocity.
Over short lag times, we can assume that swimming velocity is constant, i.e., \(\vec v_{\rm s}(t) = \vec v_{\rm s}(0) + O(\tau)\).
Thus, the displacement of the \(n\)th swimmer is approximated by
\begin{equation}
  \label{eq:swimmer-vel}
  \Delta\vec x_{\rm s}^n(\tau) = \tau \vec v_{\rm s}^n(0)
\end{equation}
where we neglect \(O(\tau^2)\) contributions.

We now introduce a ``common'' reference frame having Cartesian coordinates \((x', y')\), which places all bacteria at the origin with swimming direction in the \(+y'\) direction.
In other words, the common frame corresponds to \(N_{\rm s}\) different ``views'' of the laboratory frame, each one being co-located and co-aligned with the \(n\)th swimmer.
In this frame, we define unit vectors \(\vec e_{x'}\) and \(\vec e_{y'}\), which point in the \(x'\) and \(y'\) directions, respectively, and we may express the position vector in the common frame as \(\vec x' = x' \vec e_{x'} + y' \vec e_{y'}\).
Notice that a single point in the common frame maps to \(n\) points in the laboratory frame as \(\vec x^n = \vec x_{\rm s}^n + \tnsr Q^n \cdot \vec x' \), where
\begin{equation}
  \label{eq:Q-defn}
  \tnsr Q^n = \vec p^n \vec e_{x'} + \vec q^n \vec e_{y'}.
\end{equation}
The orthonormal pair of vectors \(\vec p^n\) and \(\vec q^n\) appearing in \eqref{eq:Q-defn} are perpendicular and parallel to the swimming direction of the \(n\)th bacterium, respectively.
Specifically, we define \(\vec q^n\) by \(\vec v_{\rm s}^n = \vec q^n v_{\rm s}^n\), where \( v_{\rm s}^n = \norm{\vec v_{\rm s}^n}\) is the swimming speed of the \(n\)th bacterium.

The covariance of the magnitude of the swimmer displacements and the displacement of a tracer particle located at a fixed position \(\vec x'\) in the common frame is given by
\begin{equation}
  \label{eq:cov-displacements}
  \vec{\chi}(\vec x', \tau)
  = \avg{\norm{\Delta \vec x_{\rm s}^n(\tau)} \, \tnsr Q^{\prime n} \cdot
    {\Delta \vec x_{\rm p}}(\vec x^n_{\rm s} + \tnsr Q^n \cdot \vec x', \tau) }_n
\end{equation}
where $\tnsr Q^{\prime n}$ is the transpose of $\tnsr Q^n$ and \(\avg{\cdot}_n\) denotes the expected value of the ensemble over all swimmers.
Considering the different contributions to \(\Delta\vec x_{\rm p}\) expressed by \eqref{eq:particle-displacement} and using \eqref{eq:swimmer-vel}, we may rewrite \eqref{eq:cov-displacements} as
\begin{equation}
  \label{eq:cov-displacement-sums}
  \begin{aligned}
    \vec{\chi}(\vec x', \tau)
    &= \tau{\avg{v_{\rm s}^n \tnsr Q^{\prime n} \cdot \sum_{m=1}^{N_{\rm s}}  \Delta \vec x_{\rm sp}^m(\vec x^n_{\rm s} + \tnsr Q^n \cdot \vec x', \tau)}}_n \\
    &\quad + \tau{\avg{v_{\rm s}^n \tnsr Q^{\prime n} \cdot \Delta\vec{x}_{\rm p,T}(\vec x^n_{\rm s} + \tnsr Q^n \cdot \vec x', \tau)}}_{n} \\
    &\quad + \tau{\avg{v_{\rm s}^n \tnsr Q^{\prime n} \cdot \sum_{m=1}^{N_{\rm p} - 1} \Delta\vec x_{\rm pp}^{m}(\vec x^n_{\rm s} + \tnsr Q^n \cdot \vec x', \tau)}}_n,
  \end{aligned}
\end{equation}
where we note that the product \(v_{\rm s}^n \tnsr Q^{\prime n}\) only depends on the swimming velocity of the \(n\)th bacterium (\(\tnsr Q^{\prime n}\) encodes swimming direction).
Regarding the first term of \eqref{eq:cov-displacement-sums}, we may assume that the bacterial concentration is sufficiently dilute that the swimming velocities of different bacteria are uncorrelated.
Then, all terms in the summation under the average are negligibly small except for when \(n = m\).
Furthermore, the second and third terms of \eqref{eq:cov-displacement-sums} can be entirely neglected.
The second term is negligible because effect of Brownian motion on the velocity of a bacterium is generally much smaller than contributions due to active swimming.
Therefore, the swimmer's displacement and the tracers' Brownian displacement are only weakly correlated.
The third term vanishes because the particle displacements due to sources other than flow disturbances produced by the bacteria are uncorrelated with the swimming velocities of the bacteria.
Moreover, the displacement of a bacterium due to interactions with tracer particles is small compared to the displacements due to active swimming.
Therefore, while \(\Delta\vec x_{\rm pp}\) is not necessarily small, its correlation with the swimmer velocity is negligible.
With these simplifications, \eqref{eq:cov-displacement-sums} reduces to
\begin{equation}
  \label{eq:cov-simple}
  \begin{aligned}
    \vec\chi(\vec x', \tau) = \tau\avg{v_{\rm s}^n \, \tnsr Q^{\prime n} \cdot \Delta \vec x_{\rm sp}^n(\vec x_{\rm s}^n + \tnsr Q^n \cdot \vec x', \tau)}_n.
  \end{aligned}
\end{equation}

Recall that \(\Delta\vec x_{\rm sp}^{n}(\vec x, \tau)\) describes the displacements of particles in the laboratory frame induced by the flow disturbance induced by the \(n\)th swimmer.
Let \(\vec u^n(\vec x)\) denote this flow disturbance.
Assuming that the tracer particles are sufficiently small, their displacement due to this flow disturbance at small lag times is given by
\begin{equation}
  \label{eq:particle-vel}
  \Delta \vec x^n_{\rm sp}(\vec x, \tau) = \tau \vec u^n(\vec x).
\end{equation}
In terms of a fixed position \(\vec x'\) in the common frame, we may express the velocity disturbance produced by the \(n\)th bacterium as
\begin{align}
  \label{eq:vel--lab<->common}
  \vec u^n(\vec x_{\rm s}^n + \tnsr Q^n \cdot \vec x') &= \tnsr Q^n \cdot \vec u^{\prime n}(\vec x').
\end{align}
Then, using \eqref{eq:particle-vel} and \eqref{eq:vel--lab<->common} in \eqref{eq:cov-simple} yields
\begin{equation}
  \label{eq:cov-velocities}
  \vec\chi(\vec x', \tau) = \tau^2 \avg{v_{\rm s}^n  \vec u^{\prime n}(\vec x')}_n,
\end{equation}
where we have used the identity \(\tnsr Q^{\prime n} \cdot \tnsr Q^{n} = \tnsr I\).

We now assume that the interface is homogeneous in its mechanics so that the velocity fields produced by the bacteria are spatially invariant.
Then, we may write \(\vec u^{\prime n}\) as the sum
\begin{equation}
  \label{eq:vel-decomposition}
  \vec u^{\prime n}(\vec x') = \bar{\vec u}'(\vec x') + \tilde{\vec u}^{\prime n}(\vec x'),
\end{equation}
where \(\bar{\vec u}' = \avg{\vec u^{\prime n}}_n\) is the mean velocity disturbance that, in the lab frame, differs from bacterium to bacterium only due to differences in their swimming directions or position.
On the other hand, \(\tilde{\vec u}^n\) represents the difference of the \(n\)th bacterium from this mean due to variations in everything other than position and direction: swimming speed, bacterium geometry, trapped state, etc.
Thus, we may interpret \(\bar{\vec u}'\) as the velocity disturbance produced by an ``average'' bacterium.

Using \eqref{eq:vel--lab<->common} and \eqref{eq:vel-decomposition} in \eqref{eq:cov-velocities}, we obtain for the swimmer-tracer displacement covariance
\begin{equation}
  \label{eq:cov--lab<->common}
  \tnsr\chi(\vec x', \tau) =
  \tau^2 \bar v_{\rm s}  \bar{\vec u}'(\vec x') +
  \tau^2 \avg{v_{\rm s}^n \tilde{\vec u}^{\prime n}(\vec x')}_n,
\end{equation}
where \(\bar v_{\rm s} = \avg{v_{\rm s}^n}_n = \tau^{-1} \avg{\sqrt{\Delta\vec{x}_{\rm s}^n \cdot \Delta\vec{x}_{\rm s}^n}}_n\) is the average swimming speed.
The second term in \eqref{eq:cov--lab<->common} can be eliminated under the assumption that swimming velocity and the corresponding flow disturbance due to a swimmer are related linearly.
This assumption is reasonable given that bacteria swim at very small Reynolds numbers.
Thus, letting \(\tilde{\vec u}^{\prime n} = v_{\rm s}^n \tilde{\vec w}^{\prime n}\), where \(\tilde{\vec w}^{\prime n}\) is dimensionless and independent of \(v_{\rm s}^n\), the second term in \eqref{eq:cov--lab<->common} reduces to
\begin{equation*}
  \tau^2 \avg{v_{\rm s}^n \tilde{\vec u}^{\prime n}}_n =
  \tau^2 \bar{v_{\rm s}}^2 \cdot \avg{\tilde{\vec w}^{\prime n}}_n,
\end{equation*}
which vanishes because \(\avg{\tilde{\vec w}^{\prime n}}_n = \avg{\tilde{\vec u}^{\prime n}}_n  / \bar v_{\rm s} = 0\).
With this term eliminated, solving \eqref{eq:cov--lab<->common} for \(\bar{\vec u}'\) yields the mean velocity disturbance as
\begin{equation}
  \label{eq:mean-vel-field}
  \bar{\vec u}'(\vec x') = \frac{\vec\chi(\vec x', \tau)}{\tau^2 \bar v_{\rm s}}.
\end{equation}

\subsection{Experimental implementation of CDV}
\label{sec:CDV-implementation}

\begin{figure}
  \centering
  \includegraphics[width=\textwidth]{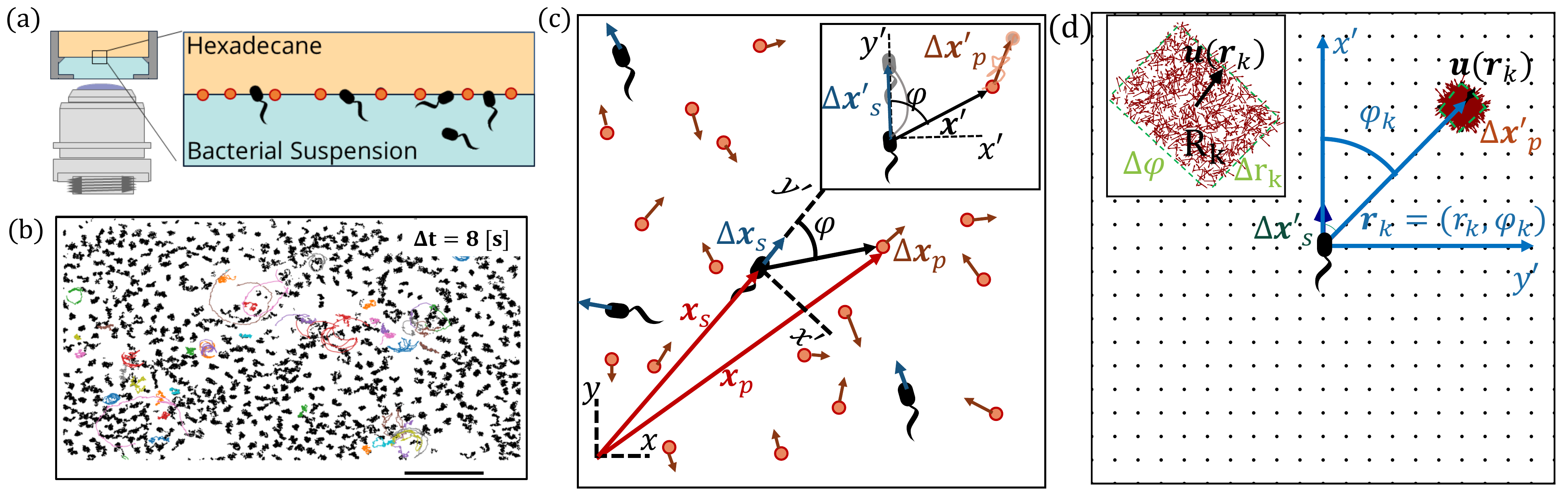}
  \caption[Implementation of Correlated Displacement Velocimetry (CDV)]{%
    \textbf{Correlated Displacement Velocimetry (CDV).} 
    Panel (a) gives a schematic of the experimental setup.
    Bacteria and passive tracers are observed on a planar hexadecane-water interface.
    Panel (b) shows trajectories of bacteria and tracers over a time interval of $\Delta t = \SI{8}{s}$; black trajectories: tracers; colored trajectories: bacteria, scale bar: \SI{20}{\micro\meter}.
    (c) Schematic of the lab frame \((x,y)\) and common frame \((x',y')\) coordinate systems: To measure the flow field from the correlated displacements of tracers and bacteria a common coordinate system is defined with each swimming bacterium at the origin and the \(y'\) axis aligned with the swimming direction. Inset: transformed displacement vectors from the lab frame to the common frame. 
    (d) To measure the flow field, tracers displacements in the common frames are sampled in different region \(R_k\) based on their locations in the common frame. Inset: zoomed in view of tracer displacements, red vectors, in the region \(R_k\) with center location of \(\vec r_k\) and size of \(\Delta r_k\) and \(\Delta \varphi\);  average displacement in the each region weighted by the displacements of the swimmers results the flow field at \(\vec r_k\).
  }
  \label{fig:CDV}
\end{figure}

To use \eqref{eq:mean-vel-field} for computing the mean bacterium-induced velocity disturbance \(\bar{\vec u}'\), we must compute \(\vec\chi\) from experimental swimmer and tracer displacement data.
We therefore evaluate particle and swimmer displacements over lag time $\tau$ at various positions $\vec{x}'$ around the swimmer directly in the common frame as illustrated in Figure~\ref{fig:CDV}(d).
As we cannot compute the true expected value represented by \(\vec\chi\), we must instead approximate \(\vec\chi\) using a sufficiently large sample size and implement a method to control sampling error.
In particular, the number of swimmer-tracer displacement products considered at each position must be large enough for the signal to emerge from Gaussian noise.
Here, our strategy is to discretize the spatial domain and to bin appropriately displacement products.

Given the form of the flow field, which decays with distance from each swimmer, it is advantageous to construct these bins using a polar coordinate system in the common frame, adjusting bin size with distance from the swimmer.
Here, the common frame acts as a ``sampling space'' where we can bin together related displacement data over multiple observations of multiple bacteria, all providing independent realizations of the ensemble.
This grid is parameterized by \(\vec x' = -\vec e_{x'} r \sin\varphi + \vec e_{y'} r\cos\varphi\), where \(r = \norm{\vec x'}\) and \(\varphi\) is the angle measured counterclockwise from the \(y'\) axis.
We then consider a set of bins arranged on this grid, where the \(k\)th bin spans the region \(R_k\) given by \(r_k - \Delta r_k / 2 < r < r_k + \Delta r_k / 2\) and \(\varphi_k - \Delta\varphi / 2 < \varphi < \varphi_k + \Delta\varphi / 2\).
We choose $\Delta\varphi$ to be the fixed value of $\pi/60$ and $\Delta r$ to increase in proportion with $r^3$.
With this discretization scheme, the number of particle displacement vectors in each bin scales as $\Delta \varphi r\Delta r \sim r^4$, ensuring a constant signal to noise ratio across the field of the measurement.
We now consider the computation of \(\vec\chi\) in each bin.
In the lab frame, the \(n\)th swimmer undergoes displacement \(\Delta \vec x_{\rm s}^n\) and the \(m\)th passive tracer undergoes displacement \(\Delta \vec x_{\rm p}^m\) over lag time \(\tau\) (Figure~\ref{fig:CDV}(c)).
In the common frame, as applied to the position and orientation of the \(n\)th bacterium, that same swimmer undergoes displacement \(\Delta\vec{x}^{\prime n}_{\rm s}\) in the \(+y'\) direction, and the passive tracer undergoes displacement \(\Delta\vec{x}^{\prime m}_{\rm p}\).
The lab-frame and common-frame swimmer and particle displacements are related by
\( \Delta \vec x_{\rm s}^{\prime n} = \vec e_{y'} \norm{\Delta \vec x_{\rm s}^n} \) and
\( \Delta \vec x_{\rm p}^{\prime m} = \tnsr Q^{\prime n} \cdot {\Delta \vec x_{\rm p}^{\prime m}}\), respectively.
Each active swimmer displacement vector \(\Delta\vec{x}_{\rm s}^{\prime n}\) is considered as a source that generates correlated displacements of tracer particles in the domain.
We then approximate \(\vec\chi(\vec x', \tau)\) for \(\vec x'\) in the \(k\)th bin as
\begin{equation}
  \label{eq:cov-displacements-binned}
  \vec\chi(\vec x' \in R_k, \tau) \approx \vec\chi^k = \avg{\norm{\Delta \vec x_{\rm s}^n} \, \Delta \vec x_{\rm p}^{\prime m}}^k_{m,n}
\end{equation}
where we use the notation \(\avg{\cdot}^k_{m,n}\) to denote the sample mean over all \(n = 1, \cdots, N_{\rm s}\) bacteria and the subset of tracer particles located in \(k\)th bin (\(\vec x_{\rm p}^{\prime m} \in R_k\)) before displacing by \(\Delta \vec x_{\rm p}^{\prime m}\).
Thus, we regard the particle index \(m\) in the average in \eqref{eq:cov-displacements-binned} as running over \(m = 1, \cdots, N^k_{\text{p,bin}}\), where \(N^k_{\text{p,bin}}\) is the number of tracer particles in the \(k\)th bin.
The mean velocity disturbance due to a bacterium is then evaluated from \eqref{eq:mean-vel-field} as
\begin{equation}
  \label{eq:mean-vel-field-binned}
  \bar{\vec u}^{\prime k} \approx \frac{\vec\chi^k}{\tau^2 \bar{v}_{\rm s}}
\end{equation}
for points lying in the \(k\)th bin.

The number of particle displacement vectors in each bin must be large enough to ensure that \(\vec\chi^k\) is a good approximation to \(\vec\chi\) for \(\vec x' \in R_k\).
This task is most difficult for points in the far field of the reported displacement field $(r > \SI{60}{\micro m})$, where displacements are weakest.
Therefore, we require $N_{\text{p,bin}}^k \ge N_\text{p,min}$, where $N_\text{p,min}$ is the minimum number of tracer particles necessary to resolve small hydrodynamic displacements $\vec{u}(\vec{r}) \tau$ in the presence of noise from random Brownian motion.
The measurement error in evaluating \eqref{eq:mean-vel-field} owing to Brownian noise in \(k\)th is proportional to
\[ \sqrt{\avg{\Delta \vec x_{\rm p}^m(\tau)^2}_{N_p}}/{2\tau \sqrt{N^k_{\rm p, bin}}}  \]
In this study, we require $N_\text{p,min}=10^6$ in the far-field.
Thus, the measurement error is orders of magnitude smaller than hydrodynamic displacements of $\SI{\sim 1}{nm}$ for particles undergoing Brownian displacements of $\SI{\sim100}{nm}$ over $\tau = \SI{0.2}{s}$.

\section{Physico-chemistry of the fluid interface}
\label{viscosity}

We study bacteria swimming at the aqueous-hexadecane interface with characteristic velocity $v= \SI{10}{\micro\meter /\second}$ and characteristic cell body size $a= \SI{1}{\micro\meter}$.
The average viscosity of the bulk aqueous and oil  phases $\bar\mu = \SI{2.17e-3}{\pascal\second}$.
The bacteria swim with negligible Reynolds number, $Re = \rho v a / \bar{\mu} \approx 10^{-4}$, where $\rho$ is the density of water.
Therefore, the fluids above and below the interface can be assumed to be in the creeping flow regime.
An independent upper bound of the surface viscosity can be determined in situ using the same CDV techniques that we use to determine the flow field around the swimming bacteria \citep{molaei_interfacial_2021}.
By analyzing the correlated motion of tracer-tracer pairs (rather than swimmer-tracer pairs as in Section~\ref{sec:CDV-theory}) via CDV, we find bounding values for the surface viscosity $\mu_s = \SI{1.5e-9}{\pascal\second\meter}$, and Boussinesq number $Bo = l_s/a < 0.7$, where $l_s = \mu_s / \bar\mu$.
We also measure the equilibrium surface tension $\sigma_{\rm eq}$ of the hexadecane-buffer interface, which we find is \(\sigma_{\rm eq} \approx \SI{46.7}{mN/m}\).
This value is less than the expected surface tension of \SI{55.2}{mN/m} \citep{goebel_interfacial_1997}.
The difference indicates the presence of adsorbed surfactant, likely from impurities in the oil, impurities in the buffer, or from bacterial secretions.

The flow from the swimming bacterium can redistribute surfactant, generating a surface pressure gradient or Marangoni stress that opposes bacteria motion.
We define the surface pressure as \(\Pi = (\sigma_{\rm eq} - \sigma)\), where \(\sigma\) is the (dynamic) surface tension.
The relative importance of the Marangoni stress to the viscous stress is captured by the Marangoni number
\begin{equation}
  \label{eq:Ma-defn}
  Ma = \frac{\bar{\Gamma}}{\bar{\mu} v} \frac{\partial\Pi}{\partial\Gamma},
\end{equation}
where $\bar{\Gamma}$ is the average surface concentration of surfactant.
In the limit of extremely low surfactant concentrations (e.g., $\bar{\Gamma}= \SI{e3}{molecules / \micro\meter^2}$), the interface exhibits a gaseous state with ${\partial\Pi}/{\partial\Gamma}\approx k_B T$.
Even in this very dilute regime, Marangoni stresses are remarkably significant; we find from \eqref{eq:Ma-defn} that $Ma \approx 150$.
Furthermore, for dilute surfactants, mass fluxes between the interface and the bulk are negligible.
The distribution of surfactants is determined by surface advection, as influenced by fluid flow disturbances produced by the bacteria, Marangoni flows, and surface diffusion.
Thus, the mass balance of the surfactant is given by
\begin{equation*}
  \vec{\nabla}_s \cdot (\vec{v}\Gamma) - D_s \vec{\nabla}_s^2 \Gamma= 0
\end{equation*}
where $D_s$ is the surface diffusivity of the surfactant, and $\vec{v}$ is the fluid velocity field at the interface.
The relative importance of surface advection versus surface diffusion is characterized by the product of Marangoni and Peclet number, $Ma\,Pe$, where Peclet number $Pe = a v / D_s$.
For typical values of $D_s$, around $\SI{10}{\micro m^2/s}$, we estimate $Pe\sim 1$.
Then, in the limit of large $Ma$, $Ma\,Pe$ is also large.
In this case, viscous stress generated by the swimmer cannot appreciably compress the surfactant monolayer, and the interface must be treated as an incompressible two-dimensional layer \citep{chisholm_driven_2021,blawzdziewicz_stokes_1999,fischer_viscous_2006}.
Similar to the to the manner in which the hydrodynamic pressure enforces fluid incompressibility in a bulk fluid, the surface pressure gradient in the interface enforces the interfacial incompressibility constraint.

\section{Characterization of the trapped state of bacteria at the interface}
\label{trap angle}
The cell bodies of the bacteria are trapped at the interface in a variety of configurations.
These configurations will affect the flow disturbance generated by each bacterium.
We estimate the configurations of the cell bodies using data accessible in our experiment.
While we do not have a side view of the cell bodies, we do record the projected image (or silhouette) of each cell body on the interfacial plane.
The cell bodies of \textit{P. aeruginosa} may be approximated as a prolate spheroid \citep{wu2015pseudomonas}. Therefore, the perimeter of each silhouette is approximated as an ellipse.
The aspect ratio of each ellipse, $\gamma$, is calculated as the length of its major axis divided by the length of its minor axis.

Weak fluctuations in $\gamma$ for each cell indicate that cell bodies do not change their configuration with time; this indicates that cells likely adsorb onto the interface with pinned contact lines. More elongated cell body aspect ratios are associated with trapped configurations that are closer to parallel to the interface. Values for $\gamma$ observed in experiment vary between a maximum value, $\gamma_{\parallel}=1.78$ and minimum value $\gamma_{\perp}=1$ for swimmers on the interface.
We define the apparent trapping angle $\theta$ for each cell body as \citep{coertjens2017adsorption}
\begin{equation}
   \label{eq:trapping-angle-def}
   \cos\theta = \frac{\gamma - 1}{\gamma_\parallel - 1}.
\end{equation}
We form ensembles of swimmers with differing $\theta$ to probe the role of the swimmers' trapped configurations in changing the balance of dominant hydrodynamic modes in the flow field and to rationalize them against theory. 

Since our experimental apparatus requires a long working distance imaging system, the numerical aperture of the objective is small leading to low diffraction limited resolution.  A conservative estimate of the uncertainty in the apparent size of the bacteria due do the diffraction limited resolution is \(\delta l \approx \lambda_w/4 NA \approx \SI{0.25}{\micro\meter}\), where \(\lambda_w\) is the center wavelength of the visible spectrum. The uncertainty in the apparent aspect ratio can be estimated as \(\delta \gamma\approx \sqrt{1+\gamma^2} \delta l/w \), where \(w\approx \SI{1.3}{\micro\meter}\) is the minor diameter of the bacteria. The uncertainty of the trapping angle is then estimated as \(\delta \cos{\theta}=\delta \gamma/(\gamma_{\parallel}-1)\). This estimate is conservative because we measure the aspect ratio of each bacterium over its entire trajectory of typical duration of \SI{10}{\second}, providing \num{\sim250} sample points per cell body, which can reduce the error. Furthermore, we obtain the cell body aspect ratio by fitting to an ellipse, using all of the data for the cell body. Finally, we find an expected relationship between the bacteria speed and $\cos\theta$.
     
We note an assumption that causes the apparent trapping angle $\theta$ to differ from the actual trapped angle $\theta^*$, to which we do not have experimental access.
The definition of $\theta$ in \eqref{eq:trapping-angle-def} assumes that the aspect ratio of cell body ${\gamma}_{\parallel}$ is identical for all bacteria, whereas bacteria cell bodies have a natural dispersion in shape.
To characterize this natural dispersion, we measure cell body aspect ratios, $\gamma_b$, of dead bacteria on a cover glass and find $\gamma_{b}=1.55 \pm 0.15$.
To estimate the difference between apparent and actual trapping angles, we relate $\theta^*$ for each cell to $\gamma$ according to
\( \gamma - 1 = \cos{\theta^*\left( \gamma_{b} - 1 \right)} \).
The error between the apparent and the actual cell body trapping angles is estimated as
\begin{equation}
    \label{eq:trapping-angle-error}
    \begin{aligned}
            \sigma_{\cos\theta^*}
             & = \left[\frac{1}{n-1}\sum_{i = 1}^{n}{\left({\cos\theta -\cos\theta^* }\right)^2}\right]^{1/2}\\
             & = \left[\frac{1}{n-1}\sum_{i = 1}^{n}{\left(\frac{\cos\theta^*\left( \gamma_{b} - \gamma_{\parallel} \right)}{\gamma_{\parallel} - 1}\right)^2}\right]^{1/2},
    \end{aligned}
\end{equation}
where \(n\) is the number of measurements of \(\gamma_{b}\) for the cell bodies on the microscope slides.
For $\gamma_{\parallel}=1.78$, \eqref{eq:trapping-angle-error} yields an error of \(\sigma_{\cos\theta^*} = 0.278 \cos\theta^*\).
Note that, according to \eqref{eq:trapping-angle-error}, the maximum error occurs when \(\theta^* = 0\) (i.e., for bacteria trapped parallel to the interface).

\section{Hydrodynamics of interfacially trapped bacteria}
\label{sec:hydrodynamic-theory}

\begin{figure*}
  \centering
  \includegraphics[width=\textwidth]{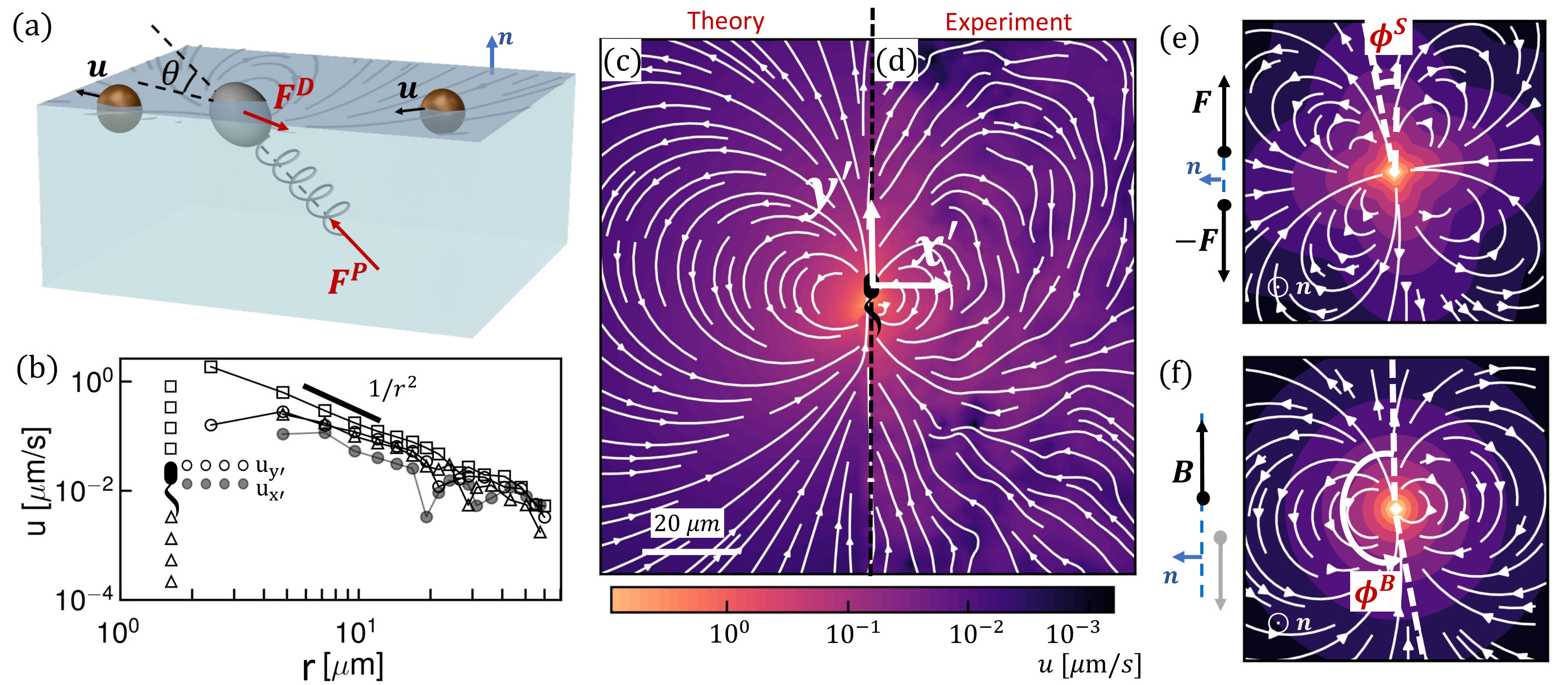}
  \caption[Interfacial flow around a pusher bacterium]{%
    \textbf{Interfacial flow around a pusher bacterium.}
    Panel (a) gives a schematic of an adsorbed bacterium and passive tracer particles that moves in the interface with propulsive force $\vec F_P$ that is resisted by drag force $\vec F_D$.
    Panel (b) plots the spatial decay of the velocity field $\vec u(\vec r)$ along the $x'$ and $y'$ axes.
    Panel (c) plots the theoretical fit of the bacterial flow field, and panel (d) plots the measured flow field generated by an ensemble of bacteria.
    Streamlines indicate the local direction of the flow, and the heatmap indicates flow speed.
    Panels (e) and (f) plot the separate \(\tnsr{S}\) mode and \(\vec{B}\) mode components of the theoretical fit in panel (c), respectively.
    These modes are superposed with the singular point of both modes shifted to position $y' = \delta y$ and have tilt angles $\phi^S$ and $\phi^B$, respectively.
  }
  \label{fig:flow-fields}
\end{figure*}

The measured flow field, shown in Figure~\ref{fig:flow-fields}(d), reveals the averaged pusher flow field generated by the ensemble of bacteria at the interface. The flow field is dominated by bacteria with speed, trajectory curvature, and apparent trapping angle near the median value of these properties (\( \bar v_{\rm s} = \SI{9.8}{\micro m/s}, \bar{\kappa} = \SI{0.2}{\micro m^{-1}},\bar{\theta} = 46^{\circ}\)).
While this flow field decays as $1/r^{2}$ in all directions (Figure~\ref{fig:flow-fields}(b)), its structure differs significantly from the fore-aft symmetric bulk stresslet generated by a pusher in the bulk. Rather, the observed flow field has broken fore-aft symmetries and distinct closed streamlines. 
The observed flow also differs significantly from the flow field reported for pushers near solid surfaces \citep{drescher_fluid_2011}. These differences are particularly notable since a viscosity-averaged stresslet with functional form like the bulk stresslet is a solution for Stokes flow in an interface characterized solely by a uniform interfacial tension \citep{chisholm_driven_2021}. 
To understand the flow structure, we consider the manner in which interfacial trapping alters swimmer hydrodynamics. Interfacially-trapped bacteria have flagella rotating in the aqueous sub-phase (Figure~\ref{fig:flow-fields}(a)) which generate the propulsion force that drives their motion. The rotation of the flagellum also generates a torque that favors rotation of the cell body in the plane of the interface. However, such rotational motion is prevented by contact line pinning. Translation perpendicular to the interface is also precluded by contact line pinning and by the interfacial tension $\sigma$, which is large compared to any viscous stresses generated by the swimmer, as described by the capillary number, $Ca=\bar{\mu}v/\sigma\approx10^{-6}$. Furthermore, the stress state of fluid interfaces is complex, with Marangoni stresses due to surface tension gradients that oppose the swimmer motion, and surface viscosities that generate dissipation.

Hydrodynamic theory for interfacially-trapped swimmers at incompressible interfaces suggests that the flow sufficiently far from the bacterium is dominated by a superposition of two flow modes, which we call the ``\(\tnsr S\)'' and ``\(\vec{B}\)'' modes \citep{chisholm_driven_2021}.
These modes are associated with different components of dipolar stress distributions exerted by the moving bacterium surface on the surrounding fluid and the interface.
The \(\tnsr{S}\) mode arises from the projection of these stresses onto the interfacial plane and is thus termed an ``interfacial stresslet''; it is analogous to the stresslet generated by Stokesian swimmers in a bulk fluid \citep{lauga_hydrodynamics_2009}.
The corresponding flow field is shown in Figure~\ref{fig:flow-fields}(e).
The \(\vec{B}\) mode, on the other hand, is associated with the protrusion of the bacterium’s body and flagellum into the fluid phases.
Finite ``vertical'' separation from the interfacial plane of viscous stresses along the bacterium surface gives rise to a finite tangential stress jump across the interface.
This stress discontinuity is supported by the gradients in surface tension that enforce the incompressibility constraint.
The flow field generated by the \(\vec{B}\) mode, shown in Figure~\ref{fig:flow-fields}(f), corresponds to the 2D potential flow due to a source-sink doublet.
The superposition of these modes leads to the asymmetric flow in Figure~\ref{fig:flow-fields}(d).

To compare experiment to theory, the strength, origin, and direction of the \(\tnsr{S}\) and \(\vec{B}\) modes are fitted to the observed flow field.
The theoretical derivation of these modes assumes that the two dipolar modes share the same origin.
Thus, to reconcile with the theory, we use the same location $y'$ for both modes but different orientations that reflect the complex distribution of forces around the cell body.
We place the origin and orientation of the two dipolar modes along the symmetry axis on bacterium’s body and weight these two modes to capture the main features of the observed flow field as shown in Figure~\ref{fig:flow-fields}c.
The analytical form of the flow field generated by the \(\tnsr{S}\) mode on the interfacial plane is given by
\begin{equation}
  \label{eq:stresslet}
  u^S_i = \frac{3 S_{jk}}{16\pi\bar{\mu}} \left[
    -\frac{3 r_ir_jr_k}{r^5} + \frac{r_i\delta_{jk}+r_j\delta_{ik}}{r^3}
  \right],
\end{equation}
where indices \(i,j,k \in \{1,2\}\), repeated indices are summed, $S_{jk}$ are the components of the second order tensor \(\tnsr{S}\), and \(r = \norm{\vec r}\).
The flow field generated by the \(\vec{B}\) mode is given by
\begin{equation}
  \label{eq:bmode}
u^B_i = \frac{B_j}{4\pi\bar{\mu}}\left[\frac{\delta_{ij}}{r^2} - \frac{2 r_ir_j}{r^4}\right],
\end{equation}
where $B_j$ is the dipolar strength vector with magnitude $B = \norm{\vec B}$.
A detailed discussion of the flow modes given by \eqref{eq:stresslet} and \eqref{eq:bmode} is provided by \citep{chisholm_driven_2021}.
The vector field of each mode is shifted vertically by $\delta y$ and rotated CCW by $\phi^S$ and $\phi^B$ to acquire the transformed vector field $\vec{u}'^S$ and $\vec{u}'^B$.
The two dimensional measured velocity field $\vec{u}(\vec{r})$ is fitted to $\vec{u}'=\vec{u}'^S +\vec{u}'^B$ using non-linear least squares fit to obtain the dipolar strength $(S, B)$, location of origins for both modes on the swimmer’s axis of symmetry $\langle 0, \delta y\rangle$ and orientations $(\phi^S,\phi^B)$.

The interfacial stresslet strength $S = \norm{\tnsr S} = \SI{0.48\pm0.05}{pN\cdot \micro m}$ is consistent with proir measurements for pusher bacteria in the bulk, and can be estimated as $S \approx F_D l$, with $F_D \sim \SI{0.2}{\pico\newton}$ and $l \sim \SI{2}{\micro\meter}$, where $F_D$ is the drag force acting on the bacterium and $l$ is the separation distance between two forces on the interfacial plane \citep{drescher_fluid_2011}.
The direction of the \(\tnsr{S}\) mode is rotated by $\phi^S = \SI{-8.0\pm2.3}{\degree}$.
The \(\vec{B}\) mode has strength $B = \norm{\vec B} = \SI{0.41\pm0.02}{\pico\newton\micro\meter}$ and is rotated by $\phi^B = \SI{192.7\pm4.0}{\degree}$ with respect to the swimming direction.
The expected strength $B\approx F_D h$, where $h$ is the distance between the location of the net thrust force and the interfacial plane. Thus, the strength of the \(\vec{B}\) mode is expected to be on the same order of magnitude of dipolar strength of the stresslet, in keeping with experiment.
The origin of the flow field reported in Figure~\ref{fig:flow-fields}(d) corresponds to the center of the cell body; in this coordinate system, the origin of both \(\tnsr{S}\) and \(\vec{B}\) modes is located slightly behind the body’s center,
displaced by $ \delta y = \SI{1.33\pm0.14}{\micro\meter}$.
This placement might be attributed to the elongated shape of flagellum.

\begin{figure}
  \centering
  \includegraphics[width=\textwidth]{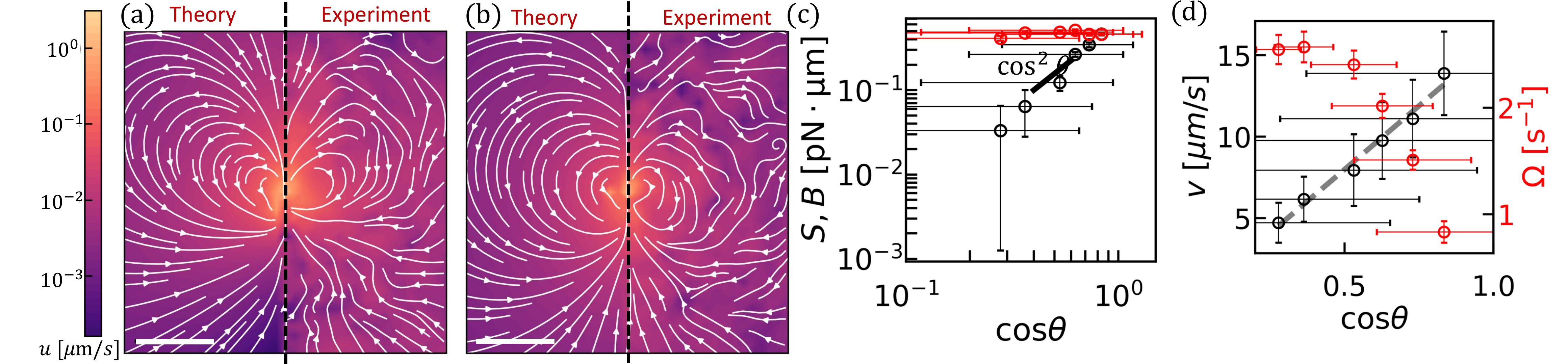}
  \caption[Flow structure depends on the trapped configuration at the interface.]{%
    \textbf{Flow structure depends on the trapped configuration at the interface.} Panels (a) and (b) are plots of the flow fields generated by bacteria with trapping angles $\theta = \SI{33\pm15}{\degree}$ (a) and $\theta = \SI{69\pm12}{\degree}$ (b).
    In each panel, prediction is shown on the left and observed fields are shown on the right.
    The heatmap and streamlines show the magnitude and direction of the flow, respectively, and the scale bar measures \SI{20}{\micro\meter}.
    Panel (c) plots the strength of the \(\vec{B}\) (red circles) and \(\tnsr{S}\) (black circles) modes versus $\cos\theta$.
    Horizontal error bars stand for the uncertainty of measuring the apparent trapping angle $\delta \cos{\theta}$.
    The vertical error bar represents the fitting error for \textbf{S} and \textbf{B} modes.
    Panel (d) plots the swimming speed (black circles) and angular speed (red circles) versus $\cos\theta$. The dashed lines are predictions based on force and torque balances, $v\approx v_{\text{max}}\cos{\theta}$.
    The vertical error bars indicate the standard error of speed and angular speed of different swimmers.}
  \label{fig:flow-fields-trap}
\end{figure}

To further analyze the dependence of the flow field generated by bacteria on their trapped configuration, we group the trajectories of swimmers with similar apparent trapping angles, calculated according to \eqref{eq:trapping-angle-def}, and measure the corresponding flow field. For the two flow fields shown in Figure~\ref{fig:flow-fields-trap}(a) and (b), the trapping angle increases from left to right. The bacteria in the more parallel configuration generate a flow field with a more asymmetric shape.
The relative strength of the \(\tnsr{S}\) and \(\vec{B}\) modes changes systematically with the trapping angle (Figure~\ref{fig:flow-fields-trap}(c)).
The strength of the \(\tnsr{S}\) mode is largest for parallel swimmers; thus, the flow fields of swimmers with small trapping angles have the most pronounced loop-like structures, strong outward flow before the swimmer, and weaker inward flow behind the swimmers, as the strong \(\tnsr{S}\) mode diminishes the inward flow driven by the \(\vec{B}\) mode. The strong dependence of the stresslet strength on trapping angle can be rationalized, noting that both $F_D$ and $l\propto\cos\theta$ so $S\sim F_D l\propto\cos ^2 \theta$.
However, the strength of the \(\vec{B}\) mode remains roughly fixed. Theory suggests that this mode has a complex dependence on the orientation and position of the thrust force, becoming weaker for forces oriented perpendicular to the interface, and stronger for larger forcing distance from the interface. These two effects give a weak dependence on trapping angle. The \(\tnsr{S}\) and \(\vec{B}\) modes are two of the dipolar modes of the stress distribution on a sphere wrapping the bacterium. Their complex dependence on trapping angle suggests that the distribution of thrust forces for bacteria with cell bodies trapped at interfaces is more complex than can be captured by a simple model of two point forces. This inability to predict this observation from this simple model may be attributed to the complex distribution of forces along the surfaces of swimmers' bodies and their flagella, which also depends on their trapped configurations at interfaces.

These changes in the balance of the two modes are accompanied by other changes in swimmer dynamics. For example, a dependence on swimming speed $v$  with $\theta$ is readily apparent; we find that $v\approx v_\text{max}\cos{\theta}$, where $v_\text{max}=\SI{15.9}{\micro m/s}$ is the maximum speed of the bacteria with the most parallel configuration (Figure~\ref{fig:flow-fields-trap}(d)).
This dependence is in keeping with a force balance on a bacterium (Figure~\ref{fig:flow-fields}(a)), which demands that the component of the propulsive force parallel to the interface $F_{P\parallel}=F_P\cos\theta$ is balanced by the drag on the swimmer $F_D=c_Dv$ where $c_D$ is the translational drag coefficient. 
 We find that the angular velocity (i.e. the product of trajectory curvature and speed) measured from swimmer trajectories $\Omega$ also varies strongly with $\theta$ (Figure~\ref{fig:flow-fields-trap}(d)), which can be rationalized by a torque balance on a bacterium at the interface.
The parallel component of the torque generated from flagellar rotation is balanced by the capillary torque.
The normal component of the torque induced by flagellar rotation $T_{P\bot}=T_P\sin\theta$ generates rotation of the bacterium perpendicular to the interfacial plane. 
In addition to this effect, the hydrodynamic torque arising from asymmetric stresses on the swimmer near the interface can also contribute to its rotational behavior, causing the parallel swimmers to swim in circular paths \citep{deng_motile_2020}. 
 
\section{Discussion}
\label{sec:discussion}

In this study, we show that the highly anisotropic environment of fluid interfaces generates unique physics that traps bacteria in the interfacial plane in diverse configurations, provides additional forms of dissipation via Marangoni stresses, and fundamentally alters the flow generated by swimmers moving in the pusher mode. This finding is significant, as fluid interfaces occur widely in nature, in medical settings, and in industrial settings. For example, air-water interfaces are encountered in puddles, lakes, and larger bodies of water, and oil-water interfaces are encountered for naturally occurring oil seeps and oil spills. Understanding the swimmers' flow fields is essential for understanding mixing, scattering and collective behaviors generated by micro-swimmers in interfacial active suspensions. In medicine, fluid interfaces are encountered in diverse organs (e.g. eyes and lungs), exposed wounds, and on medical equipment. In chemical engineering practice, fluid interfaces are widely encountered in chemical processes including in membrane reactors designed for simultaneous reaction and separation, and bubbles, emulsions, and Pickering emulsions encountered in pharmaceuticals and personal care products. While current practice has focused on passive colloids at interfaces in these settings, understanding the dynamics of active colloidal suspensions at fluid interfaces could generate new strategies to promote mixing using biomimetic active colloids.
An important feature of fluid interfaces is the juxtaposition of two fluids whose viscosities can differ from each other. Hydrodynamic theory predicts that swimming bacteria in bulk phases are hydrodynamically attracted to  fluid interfaces between phases with viscosity ratios that differ significantly from unity; this attraction may promote bacterial adsorption to the interfaces. In our prior research, we have studied the motion of bacteria at fluid interfaces with viscosity ratios of superphase:water phase $\lambda$ ranging from 0 to 300. For $\lambda\geq 20$, the majority of the interfacially trapped bacteria lose their active motion and display only Brownian displacement. However, for $0 \leq\lambda \leq 10$, the probability of swimming in curly trajectories (i.e. in pusher or puller modes) stays roughly constant, indicating that the flow fields measured here may occur at such interfaces. The measurement of swimmer's flow fields on interfaces with differing viscosity ratios is an open issue.

\begin{figure}
  \centering
  \includegraphics[width=\textwidth]{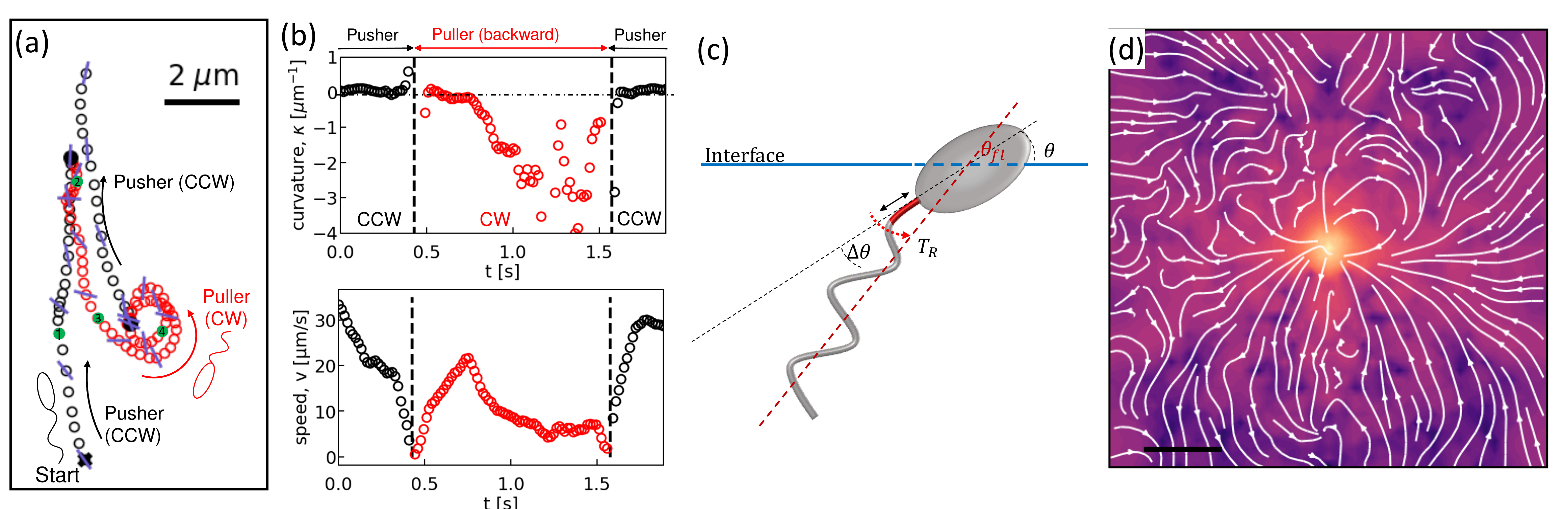}
  \caption{%
    Panel (a) shows the trajectory of a PA01 bacterium switching between CW rotation (pusher) and CCW rotation (puller) measured over a \SI{2}{s} time interval. Switching events are marked by black cross symbols.
    Panel (b) shows the trajectory curvature and the speed versus time for the bacterium trajectory shown in (a).
    Panel (c) is a schematic of an interfacially trapped bacterium in the puller mode. A reorientation torque $T_R$ due to the hydrodynamic asymmetric dipolar mode at the interface is hypothesized to bend the hook away from the interface. This effect would change the angle of the flagellum by some amount $\Delta\theta$.
    Panel (d) shows the displacement field measured via CDV around an ensemble of pullers from segments of swimmers' trajectories with CCW  curvature; the  median trajectory curvature \SI{0.38}{\micro m^{-1}}. The scale bar is \SI{10}{\micro m}.
  }
  \label{fig:puller}
\end{figure}

 We have measured the flow field around a pusher bacterium. However, unlike swimmers in the bulk, the flow field generated by a puller bacterium is not that of a pusher in reverse.  In our prior work, we have analyzed the trajectories of monotrichous bacteria that swim in pusher and puller modes and that switch between them by changing the direction of rotation of their single flagellum. Our data indicate that swimming in puller mode does not solely reverse the direction of swimming from that of a pusher, but that swimming in puller mode is accompanied by a reconfiguration of the bacterium's body configuration at the interface. In our research, we differentiate between the two modes by observing the curvature of the swimmers' paths in the interface. Bacteria swim along weakly curved CW paths in pusher mode (viewed from water side), and, after switching, move along CCW paths in the puller mode. The pusher segments of the bacteria trajectories have nearly fixed curvatures, suggesting a fixed configuration with respect to the interface. However, trajectories of bacteria moving in puller mode have curvatures that increase gradually, eventually allowing the cells to translate only weakly and to move with high rotational velocities (see Fig.\ref{fig:puller}a and Fig.\ref{fig:puller}b). We hypothesize that these features of the pullers' trajectories are related to the reorientation of the bacterial flagellum. The flagellum reorientation might result from a hydrodynamically induced torque that bends the flagellar hook and pivots the flagellum away from the interfacial plane in the puller mode, as shown schematically in Figure~\ref{fig:puller}c.

Preliminary analysis of the displacement field of tracer particles around an ensemble of pullers in a common frame indicates that this field differs significantly from that generated by pushers (see Fig.1d). In particular, the symmetries with respect to the translational axis that are present in the pusher mode are lacking in the puller mode.

\section{Conclusion}
  \label{sec:conclusion}
We measure the interfacial flow generated by a pusher bacterium, which has striking asymmetries that differ significantly from flows generated by pushers in bulk fluids. 
The flow is described by two interfacial hydrodynamic dipolar modes, an interfacial stresslet, or \(\tnsr{S}\) mode, corresponding to a parallel force dipole on the interface, and a bulk forcing or \(\vec{B}\) mode that relies on thrust from the immersed flagellum below the interface that is resisted by stresses in the interface.
This latter mode, unique to incompressible interfaces, significantly alters the flow structure, introducing the broken symmetries absent for swimmers in the bulk.
The magnitude of the \(\tnsr{S}\) mode depends strongly on the swimmer’s trapped configuration, while that of the \(\vec{B}\) mode does not.

We have focused on the flow in the plane of the fluid interface. The form of the flow field in planes adjacent to the interface or normal to the interface and their effects on mixing remain unexplored. While theory indicates that the \(\tnsr{S}\) mode generates a 3D lamellar flow with no normal component, the \(\vec{B}\) mode induces a more complex non-lamellar 3D flow. In addition, an interfacially-trapped bacterium is predicted to generate an additional pumping hydrodynamic mode with no interfacial flow that can promote mixing adjacent to the interface which is also unexplored. The implications of these flow fields in generating active interfaces to promote mixing in the interface and adjacent phases remains to be probed. 

\textbf{Acknowledgements.} This research was made possible by a grant from the National Science Foundation (NSF grant no. DMR-1607878 and CBET-1943394).

\bibliography{main}
\end{document}